# Scaling and Dissipation in the GOY shell model

**Leo Kadanoff, Detlef Lohse** *†**, Jane Wang**

The James Franck Institute, The University of Chicago,
5640 South Ellis Avenue, Chicago, IL 60637, USA

**Roberto Benzi**

Department of Physics, University of Rome "Tor Vergata",
Via della Ricerca Scientifica 1, 00133 Roma, Italy

† Correspondent, e-mail: lohse@bugle.uchicago.edu

**Abstract:**
This is a paper about multi-fractal scaling and dissipation in a shell model of turbulence, called the Gledzer-Ohkitani-Yamada model or GOY model. This set of equations describes a one dimensional cascade of energy towards higher wave vectors. When the model is chaotic, the high-wave-vector velocity is a product of roughly independent multipliers, one for each logarithmic momentum shell. The appropriate tool for studying the multifractal properties of this model is shown to be the energy flux on each shell rather than the velocity on each shell. Using this quantity, one can obtain better measurements of the deviations from Kolmogorov scaling (in the GOY dynamics) than were available up to now. These deviations are seen to depend upon the details of inertial-range structure of the model and hence are *not* universal. However, once the conserved quantities of the model are fixed to have the same scaling structure as energy and helicity, these deviations seem to depend only weakly upon the scale parameter of the model. We analyze the connection between multifractality in the velocity distribution and multifractality in the dissipation. Our arguments suggest that the connection is universal for models of this character, but the model has a different behavior from that of real turbulence. We also predict the scaling behavior of time correlations of shell-velocities, of the dissipation, and of Lyapunov indices. These scaling arguments can be carried over, with little change, to multifractal models of real turbulence.

# 1  Introduction

The recent literature contains two alternative views of the nature of well-developed turbulence. In one view, the simple scaling caught by the K41 [1] paper is the asymptotic truth which holds in the limit of high Reynolds numbers. Then the experimental facts, some of which seems to support a more complicated scaling, are described in terms of nonasymptotic corrections to scaling [2, 3, 4, 5, 6, 7, 8, 9]. In the other view [10] the experiments are better understood and described as a result of a multifractal picture [11, 12, 13] in which cascades produce anomalously large fluctuations in the velocity fields. These two views can both be supported by the experimental evidence [14, 15, 16, 17, 18, 19]. There are theoretical arguments for both.

It would be ideal to distinguish these two possibilities by direct numerical simulations of Navier-Stokes dynamics. Unfortunately, the current computing power sets an upper limit on the Reynolds number. So far, the highest Taylor Reynolds number is on the order of 200 [20, 21], which is insufficient to make the distinction. Consequently, many people have worked on simplified approaches which might offer some understanding of Navier-Stokes dynamics.

One approach, called the reduced wave vector set approximations (REWA), approximates the Navier-Stokes dynamics by representing the full velocity field on a set of wave vectors which gets more and more thinned out for higher wave vectors. For detailed discussions of the method and the results we refer to refs. [22, 23, 24, 2, 3].

Earlier, Gledzer [25] introduced another, and simpler representation of Navier-Stokes dynamics in which the velocities are placed on a one dimensional array of wave vectors. Each successive new velocity falls in a new shell in $k$-space in which the wave vector is increased by a factor of $\lambda$ so that the nth shell has $k_n = k_0 \lambda^n$. In the version of the model [26, 27, 28, 29, 30] used here and referred to as the GOY model, each shell is described by a single complex variable, the complex velocity $U_n$.

The GOY model shows at least two qualitatively different kinds of behavior [31]. In one range of parameters, the system relaxes to a time-independent state in which the velocity decays (apart from boundary corrections) with wave vector according to the predictions of K41. However, for other parameter values, the system has a long-term behavior which includes stochastic fluctuations in the velocity. The basic statistical variable is the ratio of the velocity fluctuations in neighboring shells. If this multiplier fluctuates locally, there cannot be any locking of the correlations among the fluctuations in far-distant shells. There is no way that independent short range finite strength interactions can, in one dimension, be translated into infinitely strong long ranged correlations. (The argument for this special behavior of one-dimensional systems was originally given by Landau [32] in the context of phase transition problems. Of course in higher dimensions short ranged interactions can indeed produce long-ranged correlations via phase transitions, see below.) The argument about relatively weak correlations in one dimension was first applied to the GOY model by Benzi, Biferale and Parisi [30] (hereafter referred to as BBP). They argued, whenever there are any fluctuations in the long time dynamics, the GOY model necessarily develops very considerable fluctuations in the ratio of velocities in far-distant shells. As a result, the velocities will show a multifractal behavior in the inertial range and there will be correlations among far distant shells [33, 34, 28, 30, 29, 35]. In contrast, REWA has its dynamics on a higher dimensional $k$-lattice. The dynamics can then possibly develop long range correlations among velocities with different $k$-vectors. These correlations may damp down the largest fluctuations in the velocity amplitudes and produce the observed asymptotic behavior [24, 3] which is fractal (K41) rather than multi-fractal.

Compared to REWA, the simpler GOY model offers the opportunity of a somehow easier analytic approach, and of course of less expensive numerics. REWA in turn offers the advantages of additional degrees of freedom and of an apparent extra closeness to Navier Stokes. But, we cannot tell which approach comes closer to the truth of the full Navier Stokes system.

In this paper we aim for the more modest goal of finding some of the properties of the GOY model. Specifically we wish to

1. Develop a more accurate numerical method of determining the scaling exponents of the multifractal spectrum and thereby to better understand the spectrum of fluctuations of $|U_n|$.

2. Understand the role of conserved quantities.

3. Examine whether the multifractal properties dependent upon the exact form of the dissipation and of the nonlinear term.

4. Describe and verify a theory for time dependence of various quantities of the dissipation and thereby generate a scaling theory for the multifractal properties of the dissipation, for velocity correlations, and for fluctuations in Lyapunov indices.

In the next section of this paper, we describe the GOY model and some of its major qualitative features. Section three is devoted to the properties of the velocity in the inertial range. We show how to get more accurate values of the multifractal scaling exponents than were available up to now. We then see how the exponents depend upon the various parameters of the model. In section four, we do some calculations related to time-dependent correlations in the model: subsection one describes the relation between the multifractal properties of the dissipation and that of the velocity in the inertial range; the next estimates the order of magnitude of velocity correlations; and the last discusses fluctuations in the Lyapunov index for the model.

# 2 The GOY model of turbulence

The GOY model describes a one-dimensional cascade of energies among a set of complex velocities, $U_n$, on a one dimensional set of wave vectors

$$k_n = k_0 \lambda^n, \qquad n = 1, 2, ..., N. \tag{1}$$

The model is a system of ODE with the following structure:

$$\frac{d}{dt} U_n = -D_n + F_n + iC_n. \tag{2}$$

Here $D_n$ stands for a dissipation term, $F_n$ stands for a forcing term which is only set on low-$n$ shells, and $C_n$ stands for nonlinear couplings among different shells. The last term is crucial to inducing energy cascades in the model.

We shall make the same choices for the three terms as have been used in several previous turbulence studies[25, 26, 27, 28, 30, 29, 31] :

$$\begin{align}
D_n &= \nu k_n^2 U_n, \tag{3} \\
F_n &= f \delta_{n,4}, \tag{4} \\
C_n &= a k_n U_{n+1}^\star U_{n+2}^\star + b k_{n-1} U_{n-1}^\star U_{n+1}^\star + c k_{n-2} U_{n-1}^\star U_{n-2}^\star. \tag{5}
\end{align}$$

They are intended to capture some of the features in hydrodynamics: viscous dissipation [1] of energy, external forcing on a large scale, and quadratic interaction among different modes with strength proportional to $k$. Furthermore, we impose rigid boundary conditions on $U_n$ in which the only non-zero $U_n$'s are those for which $n$ is within the range $[1, N]$. The constants, $a$, $b$, $c$, $f$, $\lambda$, and $k_0$ define the model. Throughout this paper we make the conventional choices

$$\begin{align}
k_0 &= 2^{-4}, \tag{6a} \\
f &= 5(1+i) \cdot 10^{-3}, \tag{6b} \\
a &= 1, \tag{6c} \\
b &= -\epsilon, \tag{6d} \\
c &= -1 + \epsilon. \tag{6e}
\end{align}$$

The standard case which we will use for comparison with the results in the literature has

$$\begin{align}
\lambda &= 2, \tag{7a} \\
\nu &= 10^{-7}, \tag{7b} \\
\epsilon &= 1/2, \tag{7c} \\
N &= 22. \tag{7d}
\end{align}$$

---

[1] Later on, we will also allow for hyperviscosity $D_n = \nu k_n^\phi U_n$ and study the behavior at $\phi = 4, 6$.

Such a system is similar to the 3d Navier-Stokes dynamics in four respects:

1. In the inviscid and non-forcing limit, there are two conserved quantities [25] which can be identified with the total energy $\int |\boldsymbol{u}(\boldsymbol{x})|^2 d\boldsymbol{x}/2$ and the helicity $\int \boldsymbol{u} \cdot \boldsymbol{\nabla} \times \boldsymbol{u} d\boldsymbol{x}$ of the exact problem.

2. The cascade term conserves the phase volume, defined as the total volume in the $N$-dimensional complex velocity space. The result is a direct consequence of the statement that
$$\frac{\partial}{\partial U_n} C_n = 0. \tag{8}$$

3. The system can reach a steady state in which it behaves chaotically. Since the system is forced at large scale and the dissipation occurs mostly at small scales, the system must cascade energy from large scales to smaller ones.

4. The multifractal behavior shows some resemblance to the behavior seen experimentally.

A general discussion of conserved quantities can be found in Gledzer's work [25]. We simply notice that the cascade terms give rises to the expression for the general conserved quantity:
$$W = \sum_n |U_n|^2 z^n, \tag{9}$$
whenever z satisfies the quadratic equation
$$0 = a + bz + cz^2. \tag{10}$$

We require one of the two conserved quantities to have the structure of the kinetic energy, i.e. $z = 1$:
$$E = \sum_n |U_n|^2/2. \tag{11}$$

Then from equation (10), we must take $a + b + c = 0$, as reflected in equations (6c) - (6e). By adjusting the time scale, we can make $a = 1$. So the cascade terms of the model contain only two free parameters, which can be defined to be $\epsilon$ and $\lambda$. Now the energy conservation law takes the form
$$\frac{d}{dt}|U_n|^2 = -\nu k_n^2 |U_n|^2 + \Re[f U_n^\star \delta_{n,4}] + J_{n-1} - J_n. \tag{12}$$

Here the different terms refer respectively to the dissipation, to the forcing, and to a discrete divergence of an energy flux. The cascade produces the fluxes, which are defined in terms of the triple products:
$$\Delta_n = k_{n-1} U_{n-1} U_n U_{n+1}. \tag{13}$$

Then the energy flux from the $n$th mode to the $n+1$th mode is
$$J_n = \Im[-\Delta_{n+1} - (1-\epsilon)\Delta_n]. \tag{14}$$

We therefore can picture a steady state of the dynamical system as a cascade of energy from large "eddies" to smaller ones, where the energy is dissipated through viscous diffusion. It is in this sense that we say the dynamics may mimic real turbulence.

Given an $\epsilon$, the other conserved quantity is then determined by equation (9) in terms of the other solution of equation (10), $z = 1/(\epsilon - 1)$ as

$$H = \sum_n |U_n|^2 (\epsilon - 1)^{-n}. \tag{15}$$

The corresponding conservation law takes the form of

$$\frac{d}{dt}(\epsilon - 1)^{-n}|U_n|^2 = -\nu k_n^2 |U_n|^2 (\epsilon - 1)^{-n} + \Re[fU_n^\star \delta_{n,4}](\epsilon - 1)^{-n} + L_{n-1} - L_n, \tag{16}$$

where,

$$L_n = (\epsilon - 1)^{-n} \Im[(\Delta_n - \Delta_{n+1})] \tag{17}$$

is the relevant flux from the $(n-1)th$ mode to the $nth$ mode.

The inviscid three dimensional hydrodynamics has two conserved quantities quadratic in velocity: the total energy and the helicity; the latter is a spatial integral of velocity dotted into vorticity. In our shell model, the closest analog to helicity is $\sum_n (-1)^n k_n |U_n|^2$. Whenever $1/(1-\epsilon)$ is equal to $\lambda$, the helicity defined in this way agrees with our second conserved quantity $H$. In particular, this equality holds for the conventional choice of parameters (6) with $\epsilon = 0.5$.

We remark that the second conserved quantity plays an essential role in the theories [37, 3, 4, 38] which use corrections to scaling to explain why K41 does not fully fit experimental turbulence data. In their approach, the leading correction to scaling is the result of an additive correction to the asymptotic scaling.

The structure of the cascade part of the equation of motion is particularly interesting. This part has the form

$$C_n = \sum_{m,p} J_{n,m,p} \frac{\delta E}{\delta U_m} \frac{\delta H}{\delta U_p}. \tag{18}$$

Here $E$ and $H$ are our two conserved quantities and $J$ is proportional to a completely antisymmetric function of its three indices. The equation of motion generated from equation (18)

$$\frac{d}{dt} U_n = \{U_n, E\} \quad \text{with} \tag{19a}$$

$$\{A, B\} = i \sum_{n,m,p} J_{n,m,p} \frac{\delta A}{\delta U_n} \frac{\delta B}{\delta U_m} \frac{\delta H}{\delta U_p} + \text{c.c.} \tag{19b}$$

looks as if it might be Hamiltonian in character in that it is generated by an antisymmetrical bracket structure. The form given by equation (19b) suggests that what we are seeing is possibly a Lie-Poisson System [39]. However, detailed investigations by one of us (MM) have shown that this bracket in particular and any other possible Lie-Poisson-like bracket generating the equations of motion fails to satisfy the Jacobi identity, so that the brackets in question are not really Poisson brackets.

The numerical integration of the shell equation (2) was done using the 'lsode'-software package [40], a differential equation solver that handles the stiffness of the equations efficiently. We benchmark our code by running a well studied case of equation(7) [29, 28, 26], in which the total number of shells is 22. Our results agree with the previous ones. The satisfaction of the balance eqs. (12) and (16) for the stationary state provides an extra check for our numerics.

# 3 Wiggles in the average of $|U_n|$

## 3.1 Static Solutions

Biferale *et al.* [31] studied two classes of solutions to the GOY model. They observed numerically (for $\lambda = 2$ and $\nu = 10^{-7}$) that when $0 < \epsilon \leq 0.3843$, the system reaches a stable static solution of the Kolmogorov type. On the other hand, when $0.3953 \leq \epsilon \leq 2$ their analysis shows a chaotic and time-dependent behavior. Much of the previous analysis of the GOY model has been done for $\epsilon = 0.5$. Since the phase transition at $\epsilon \sim 0.4$ apparently has the character of a continuous transition, and since $\epsilon = 0.5$ is reasonably close to $0.4$, we should expect many of the qualitative features of the static state to manifest themselves for $\epsilon = 0.5$.

To understand the structure of the nonlinear interaction, Biferale, *et al.*[31] analyzed the static solutions in the inviscid and homogeneous GOY model to obtain an iteration map for the ratios of velocity. Here we offer a slightly different approach which gives an analytic solution for the ratios in the inhomogeneous and inviscid case. Let us first consider the case in which $\nu = 0$ and $f = 0$. For $n > 4$, one sets the cascade operator of equation (5) equal to zero and finds a linear difference equation for the product of three velocities, $\Delta_n$, defined by equation (13):

$$\Delta_{n+1} - \epsilon \Delta_n - (1 - \epsilon)\Delta_{n-1} = 0. \tag{20}$$

This equation admits a general solution of the form:

$$\Delta_n = A + B(\epsilon - 1)^n, \tag{21}$$

where $A$ and $B$ are each complex constants, related to the fluxes for energy and the other conserved quantity by

$$J_n = \Im[(\epsilon - 2)A] = \Re(fU_4^*) \tag{22a}$$
$$L_n = \Im[(\epsilon - 2)B] = \Re(fU_4^*)(\epsilon - 1)^{-4} \tag{22b}$$

Notice that, in the static case, the ratios of $U_n$'s can be expressed in terms of the ratios of $\Delta_n$'s, in particular,

$$r_n \equiv U_{n+3}/U_n = \Delta_{n+2}/(\lambda\Delta_{n+1}). \tag{23}$$

Rewriting equation (20) in terms of $r_n$ gives the ratio map obtained by Biferale et. al.. The general expression (21) permits two simple scaling solutions corresponding to $B = 0$ and $A = 0$, respectively. However, if $A$ and $B$ are both of order unity, then, as long as $0 < \epsilon < 2$, the first scaling solution ($B = 0$) will always dominate for large $n$.

For the case $B = 0$, $A \neq 0$, we have $\Delta_n = k_{n-1}U_{n-1}U_nU_{n+1} = A$, independent of $n$, and $r_n = 1/\lambda$. Note that the velocity amplitudes show period three oscillations superimposed on the classical $k_n^{-1/3}$ falloff, i.e.

$$U_n = \begin{cases} u_0 k_n^{-1/3} & \text{for } n = 0 \pmod 3 \\ u_1 k_n^{-1/3} & \text{for } n = 1 \pmod 3 \\ u_2 k_n^{-1/3} & \text{for } n = 2 \pmod 3 \end{cases} \tag{24}$$

The period three oscillations will be important in what follows.

For the second scaling solution ($A = 0$, $B \neq 0$) we have $r_n = (\epsilon - 1)/\lambda$. In particular for $\lambda = 1/(1 - \epsilon)$ (e.g., $\lambda = 2$, $\epsilon = 1/2$), where $L_n$ can be considered as helicity flux, this results in $|U_n| \propto k_n^{-2/3}$ [37, 3, 4], again, up to a period 3.

Let us now allow for the inhomogenity in the difference eqs. (20) by including the forcing term $fU_n^*\delta_{n4}$, but still neglecting viscosity. In the limit as viscosity goes to zero, $U_3$ also goes to zero. Consequently, for the time-independent situation, we find $\Delta_4 = 0$ and then, $A/B = (\epsilon - 1)^{-4}$. In this way, we derive

$$r_n = \frac{1 - (\epsilon - 1)^{n-2}}{\lambda[1 - (\epsilon - 1)^{n-3}]}, \qquad n > 3. \tag{25}$$

This expression for $r_n$ has an oscillatory behavior for $n$ in the inertial range whenever $\epsilon < 1$. Note that this solution is determined by our choices of boundary conditions and the stirring mechanism.

In table 1 we show the comparison of the numerical results and the analytical expression (25) for $r_n$ in the inertial range when $\epsilon = 0.3$ (static solution) and $\epsilon = 0.5$ (chaotic solution where we define $r_n = \langle U_{n+3}\rangle/\langle U_n\rangle$), respectively. In the former case, the agreement between the numerics and the theory is very good, with small deviations resulting from neglecting the viscous term in (25). In the latter case, we notice that although the values do not match closely, nevertheless the static theory (from which we derived (25)) describes correctly the oscillatory behavior of $r_n$. This suggests that some of the quantities in the static solutions may well reflect the properties of chaotic solutions.

## 3.2 Velocities, Triple products, and Fluxes

Previous works [28, 29, 30] studied moments of the $U_n$'s:

$$s_{n,q} = \langle |U_n|^q \rangle \propto k_n^{-\zeta_q}. \tag{26}$$

Numerically, we observe that the scaling of the $U_n$'s exhibits, superimposed on the overall power law scaling, the period three oscillations which have been reported earlier by Pisarenko et. al.[29]. The phenomenon is reminiscent of and probably has the same origin (cf. eq. (24)) as the period three oscillations in the static Kolmogorov state. These oscillations are shown in Figure 1 for two cases with different forms of dissipation: a normal viscosity case ($D_n = \nu k_n^2 U_n$) vs. a hyperviscosity case ($D_n = \nu k_n^4 U_n$). Oscillations are present in both cases, but much more pronounced in the latter situation.

We could improve our scaling analysis if we did not have the period three oscillation which confuses the pictures of power law decay. In the static case a product of the form $(s_{n-1,q} s_{n,q} s_{n+1,q})^{1/3}$ would not show any oscillation. To verify that this result extends to the chaotic situation, we plotted $(s_{n-1,q} s_{n,q} s_{n+1,q})^{1/3}$. Indeed, the oscillations disappear.

The period three oscillations are responsible for the large uncertainties in measuring the scaling exponents. This uncertainty can be eliminated by extending our analysis of the static state in the previous section. We first observe that relations similar to (20) (21) (22b) hold for $\Im \langle \Delta_n \rangle$, and we obtain:

$$\Im \langle \Delta_n \rangle = k_{n-1} \Im \langle U_{n-1} U_n U_{n+1} \rangle = -\frac{\Re \langle f U_4^* \rangle}{2 - \epsilon} \left( 1 - (\epsilon - 1)^{n-4} \right), \qquad (27)$$

We are then motivated to study the scaling of the triple moments

$$S_{n,q} = \langle |\Im[U_{n-1} U_n U_{n+1}]|^{q/3} \rangle \propto k_n^{-\zeta_q}, \qquad (28)$$

which are free of period three oscillations. As one can see from Figure 2 these quantities do show a cleaner scaling form than the $s_{n,q}$.

However, the oscillations at small $n$ still remain a problem. We then make use of the analog to the Kolmogorov structure equation [33]

$$\langle J_n \rangle = -\Im \left( \langle \Delta_{n+1} \rangle + (1 - \epsilon) \langle \Delta_n \rangle \right) = \Re \langle f U_4^* \rangle. \qquad (29)$$

In the inertial range, $J_n$ should be independent of $n$. Eq. (29) suggests that we could eliminate small-n oscillations by studying the scaling of

$$\Sigma_{n,q} = \langle \left| \Im \left[ U_n U_{n+1} U_{n+2} + \left( \frac{1-\epsilon}{\lambda} \right) U_{n-1} U_n U_{n+1} \right] \right|^{q/3} \rangle, \qquad (30)$$

which, in mean field approximation, scales as

$$\Sigma_{n,q} \propto k_n^{-q/3}. \qquad (31)$$

Expression (31) has the same scaling behavior as $S_{n,q}$, but is free from the oscillation for small $n$ in the inertial range.

This sequence of operations enable us to reduce the statistical error in $\delta \zeta_q$ from 0.05 down to 0.005, and to increase the scaling region by perhaps 1 or 2 decades.

The improved accuracy allows us to study the dependence of scaling exponents on various parameters in the model such as $\epsilon$, and the form of dissipation [2].

Note that the comparably large spectral strength $s_n$ for $n = 17$ (cf. fig. 1), i.e., just before the viscous range sets in, can be understood as a kind of bottleneck phenomenon [5, 6]. As pointed out in ref. [6], the bottleneck energy pileup is more pronounced for the hyperviscous case, since viscosity sets in more rapidly. On the other hand, there cannot be a bottleneck energy pileup for the quantities $S_n$ and $\Sigma_n$, because these quantities are based on the energy flux, which is constant in the inertial range.

## 3.3 Multiscaling

We have shown that by analyzing first $s_{n,q}$, then $S_{n,q}$, and finally $\Sigma_{n,q}$, the accuracy of the scaling exponents $\zeta_q$ can be improved step by step[3]. Besides the statistical errors reported above, the analytical expressions (27) and (29), which state $\zeta_3 = 1$, help us to control the systematic error in the $\zeta_q$ determination, since consequently $S_{n,3}$ and $\Sigma_{n,3}$ should scale as $\propto k_n^{-1}$. The numerical value for the scaling exponent of $\Sigma_{n,3}$ is indeed very close to 1, e.g., $\zeta_3 = 1.005$. We choose the averaging time for each simulation by demanding that $s_{n,q}$, $S_{n,q}$, and $\Sigma_{n,q}$ have all settled down to their long-term value. This stability is ensured by comparing successive runs, for each of which the averaging time of 15 000 large eddy turnover times turns out to be more than sufficient for $N = 22$. The inertial range is determined by the condition that $\zeta_3$ is very close to 1. The numerical values for the deviations from Kolmogorov scaling $\delta\zeta_q = \zeta_q - q/3$ presented below are measured from $\Sigma_{n,q}$.

In order to compare the $\delta\zeta_q$ for different parameters and to make claims on what they depend it is rather important to perform a very careful analysis of possible errors of our determinations. Besides unknown systematic errors there are three controllable source of errors: (i) The inertial range (and thus the fitting range for the straight line fit) is not unambiguously determined by the condition "$\zeta_3$ very close to 1". In practice, $\zeta_3$ is typically off from 1 by an amount of order $\pm 0.005$ for typical fits in the n-ranges from 4 to 15 or 5 to 15. We attempt to minimize this source of error by dividing all measured $\zeta_q$ by $\zeta_3$. Nevertheless, the $\zeta_q$ slightly depend on the fitting range. (ii) There is some statistical error from the least square fit of the data to a straight line. (iii) Results may differ from run to run, though each is over 15 000 to 50 000 large eddy turnover times.

The errors from (i) and (ii) are about of the same size, whereas that from (iii) is clearly smaller. So, with error propagation, the total standard deviation is about 1.5 times as large as the statistical error (ii). These are the errors given in figs. 3-5.

To visualize the differences in the $\delta\zeta_q$ for different sets of parameters, we found it convenient to plot $\delta\zeta_q/(q(q-3))$ rather than $\delta\zeta_q$ itself. The reason is that we want to eliminate the trivial agreement of the $\delta\zeta_q$ for $q=3$ and $q=0$, i.e., $\delta\zeta_3 = \delta\zeta_0 = 0$ for all kinds of parameter sets.

---

[2]Note that Biferale et al. [31] eliminated the period 2 oscillations by changing the boundary conditions of (2).

[3]After completing much of this work, we were told by Z.S. She that his group had independently derived a similar method for analysis of the multifractal properties of the GOY model.

Figure 3 shows the $\delta\zeta_q$ and their errors – calculated and displayed as described above – for the standard GOY model parameters, $\lambda = 2$, $\epsilon = 1/2$. In the same figure we also give the $\delta\zeta_q$ obtained from She and Leveque's (SL) [41] result

$$\zeta_q = q/9 + 2\left[1 - \left(\frac{2}{3}\right)^{q/3}\right], \qquad (32)$$

which is picked to be a simple phenomenological description of a situation in which one has an upper cutoff to the size of the turbulent fluctuations. Note that in any (finite time) numerical calculation such an upper bound will be given. Their result is in good agreement with experiment. For $q > 3$ we find surprisingly good agreement also with the numerical GOY results at the standard parameter value. This means that the *tails* of the velocity probability distribution function (PDF), corresponding to *large* fluctuations, are well described by the SL theory. For $q < 3$ we find slightly worse agreement, i.e., the *small* fluctuations (peak of PDF) are not so well described by SL.

However, the GOY-model values of $\delta\zeta_q$ are not universal. They depend on the choice of $\epsilon$ and $\lambda$. For example, for $\epsilon = 0.3$ and $\lambda = 2$, $\delta\zeta_q$ shows classical Komolgorov scaling, see Biferale et. al.[31]. Our numerical analysis shows that even when we are in a chaotic situation, $\delta\zeta_q$ varies with $\lambda$ and $\epsilon$. As seen in figure 3, e.g. for $\epsilon = 0.7$, $\lambda = 2$ the scaling corrections are much larger than for the standard case. So why should the parameter values $\lambda = 2, \epsilon = 0.5$ be so special as to agree with She and Leveque? We recall our remark made in section 2 about the second conserved quantity. As we have pointed out, given the above special parameters, the extra conserved quantity closely resembles *helicity*, which is indeed a conserved quantity in 3D inviscid hydrodynamics. We also believe that specific forms of conservation laws will have crucial influence on the outcome of statistical averages of the dynamical quantities. Consequently, we hypothesize that the scaling corrections should be invariant along a curve in $\epsilon - \lambda$ plane, which conserves both energy and helicity. The curve defined by the above constraints takes the form

$$\lambda = \frac{1}{1 - \epsilon}. \qquad (33)$$

To test this hypothesis, we experiment numerically with systems of different $(\epsilon, \lambda)$. In Figure 4 we present the scaling corrections $\delta\zeta_q$ for three pairs of $\epsilon$ and $\lambda$ 'on the curve' (33). And as a comparison, we show a case for which the parameter-values are 'off the curve': $(\epsilon = 0.7, \lambda = 2)$. The cases 'on the curve' are grouped closely together, while the case 'off the curve' is much further away. Our numerical findings suggest that the scale factor $\lambda$ plays a relatively minor role in the dynamics *once* the scaling for the second conserved quantity is picked (via equation(33)) to correspond to scaling for the helicity. Returning to our original question, we are ready to claim that the canonical choice of ($\epsilon = 0.5, \lambda = 2$) is a special one because it lies on the energy-helicity curve (i.e., respects both the energy and helicity as conserved quantities), but it is not a unique one because there are many other parameters on the same curve which will give roughly the same values of $\delta\zeta_q$.

After having compared that curve with the She-Leveque theory [41], it is also interesting to compare it with the approximations described in BBP. There are two steps of approximation in the that paper. In the first, described in section 3 of BBP, an approximate recursion relation relating different $\delta\zeta_q$ is derived. These recursions can be solved if one knows the results at $q = 1$ and $q = 2$. Using these values as adjustable parameters, we have derived all the integer-index $\delta\zeta_q$. The parameters can be adjusted so that these results agree within our numerical errors with the GOY-model numerical solutions. Moveover, in section 4 of BBP, their theory is extended to give an approximate value of all $\delta\zeta_q$ from first principles. When we check this part of their theory, we find that there are severe differences between their solutions and the numerical calculations. We thus conclude that while the section 3 results might be accurate, the section 4 results disagree with the facts.

As suggested in the conclusion of BBP, correlations in the multiplicative process among the shells may play an important role in order to obtain an accurate theoretical estimates of the scaling properties using the closure equations.

## 3.4 Dependence upon the form of dissipation

A further interesting issue is whether the scaling corrections $\delta\zeta_q$ – besides depending on $\lambda$ and $\epsilon$ – also depend on the magnitude and form of the the viscous damping term. In particular, we take the damping term to have the form $D_n = \nu k_n^\phi U_n$ and then focus upon the issue of whether the scaling results depend upon $\nu$ and $\phi$. The analysis of BBP [30] suggests that $\delta\zeta_q$ is determined only by the form of the cascade term, since it is determined by short-ranged correlations between the different shells. She [42] suggested an alternative viewpoint: Energy fluctuations are produced at all length scales and tend to cascade downward toward smaller lengths or higher $n$-values. When they enter the viscous subrange, they see a changed environment, because the viscosity term then effectively enters the equation of motion. Depending upon the value and form of the viscosity term, more or less energy might be reflected toward lower $n$-values. Thus the energy would flow through partially a direct and partially an inverse cascade. The amount of reflection would determine the corrections to scaling indices. The details of the reflection will be determined in part by the form and magnitude of viscous damping.

Let us look at the facts once again by plotting $\delta\zeta_q/(q(q-3))$ as a function of $q$ for various cases. Now we fix $\lambda$ and $\epsilon$ to their standard values and vary $\nu$ and $\phi$. We determine the $\delta\zeta_q$ and their errors exactly as pointed out in the previous subsection. The various results for the deviations $\delta\zeta_q$ – though slightly different – are reasonably close together, cf. figure 5. The results for different types of viscosity differ by about two standard statistical errors.

We have also done a preliminary analysis of the $\nu$-dependence. In the hyperviscous case the results seem to be rather somewhat dependent on $\nu$. For $s = 6$, we found, for different values of $\nu$, both stronger and weaker inertial range scaling corrections than for the standard $s = 2$ situation.

Our numerical analyses seems to hint that $\delta\zeta_q$ might depend on the type of viscosity. However, our result is certainly not accurate enough to be definitive. Many previous workers have gotten confused by crossover effects or by corrections to scaling. However, it is to be expected that future workers will obtain improved results, perhaps sufficient to tie down the errors and settle the issue.

# 4 Time correlations

## 4.1 Connection between multifractal behavior of $U_n$ and dissipation

In this section, we shall move back and forth between the shell model and the theory of real turbulence. In particular, we shall wish to contrast the two theories of multifractality, which bear many resemblances. However, as we shall see there is also a crucial difference, which is rooted in the nature of the two situations. In real turbulence, if there is a large scale velocity $\boldsymbol{U}_0$, then any small scale disturbance will produce a time derivative of the velocity

$$\frac{\partial U(\boldsymbol{R}, t)}{\partial t} = -\boldsymbol{U}_0 \cdot \nabla \boldsymbol{U}(\boldsymbol{R}, t). \qquad (34)$$

This time derivative produced by spatial variation and carried by the large-scale motion is called *sweeping*. There is no analog of sweeping in the GOY model since there is no coupling between $U_0$ and the large-$n$ $U_n(t)$. This distinction will produce a considerable difference in the answer which we will obtain for the multifractal properties of dissipation.

Recall that in the multifractal approach to the usual turbulence theory [11], averages of the velocity difference at a distance r have the form

$$< [U(\boldsymbol{R}+\boldsymbol{r}, t) - U(\boldsymbol{R}, t)]^q > \sim V_o^q [r/L]^{\zeta_q}, \qquad (35)$$

where $V_0$ is a typical value of the velocity, $L$ is an integral scale and $\zeta_q$ is the multifractal scaling exponent for the velocity. The $< \cdots >$ in equation (35) stands for an average over space and time. The analogous quantity within the shell model is $s_{n,q}$ (or $S_{n,q}$ and $\Sigma_{n,q}$ correspondingly), which scales with wave vector $k_n$,

$$s_{n,q} \sim k_n^{-\zeta_q}, \qquad (36)$$

where now the average implied is a time average. In both cases, the deviations of $\zeta_q$ from $q/3$ measure the deviation from the K41 description.

Next we look at another form of multifractal behavior. As the energy cascades towards higher values of $n$ it reaches sufficiently high wave vectors so that the viscosity can produce energy dissipation. According to equation (12) the rate of energy dissipation is [4]

$$\epsilon(t) = \nu \sum_n k_n{}^2 |U_n(t)|^2. \tag{37}$$

An analogous dissipation occurs in the Navier Stokes theory. There the dissipation depends upon both space and time and takes the form

$$\epsilon(\boldsymbol{R}, t) = \frac{1}{2}\nu \sum_{i,j}(\partial_i u_j(\boldsymbol{R}) + \partial_j u_i(\boldsymbol{R}))^2. \tag{38}$$

To get an appropriate fluctuating quantity we take the dissipation and average it over a time interval $\tau$. For the GOY case take the average to be

$$\epsilon_\tau(t) = \int_t^{t+\tau} \epsilon(t')dt'. \tag{39}$$

In contrast, Meneveau and Sreenivasan [43] looked at real turbulence data obtained by taking a particular point in space and averaging over a period of time. They measured time averages as

$$\epsilon_\tau(t) = \int_t^{t+\tau} \epsilon(\boldsymbol{R}, t')dt'. \tag{40}$$

To be more precise, $\epsilon(\boldsymbol{R}, t)$ was substituted by its one dimensional surrogate $\epsilon'(t) \sim (\partial_t u_1)^2$. The dissipation is thus measured always at the same position and there is no reason to include its $\boldsymbol{R}$-dependence. In both situations, the multifractal behavior of the dissipation is defined by considering averages of powers of the dissipation in the form

$$< [\epsilon_\tau(t)]^q > \sim \tau^{\mu_q}. \tag{41}$$

If $\mu_q$ is not proportional to $q$ for $\tau$ in the inertial range, then the dissipation is said to be multifractal.

Following [10] [11] [48], Benzi et al [36] developed a theory of this multifractal dissipation for the case of Navier Stokes turbulence, noticing that the dissipation of energy was fed through a cascade in which the flux goes through all scales of $r$ up to and including the dissipation scale.

The scales in space and time are connected by the sweeping process. If the inertial scale velocity is of order $U_o$ then the scales are connected by Taylor's hypothesis [45]

$$\tau \sim \frac{r}{U_0} \tag{42}$$

---

[4] The reader should not mix up the energy dissipation $\epsilon(t)$ with the GOY model parameter $\epsilon$ introduced in eq. (6d). We shall not refer to the latter in this section of the paper.

According to the analysis of [44, 10, 43, 11], $\mu_q$ is completely determined by $\zeta_q$. This analysis says that to an order of magnitude, the dissipation on scale $r$ is set by the flux at that scale, which can be estimated as

$$J_r(t) \sim r^{-1} U_r^3(t), \tag{43}$$

where $U_r$ is the velocity difference on scale $r$. Putting together equations (41), (42), and (43) one finds that the dissipation on scale $\tau$ has the order of magnitude

$$\langle \epsilon_\tau^q \rangle \sim\, < r^{-q} [U_r(t)]^{3q} > \sim \tau^{-q+\zeta_{3q}}, \tag{44}$$

concluding thus that $\mu_q$ is determined by the multifractal scaling of velocity according to

$$\mu_q = -q + \zeta_{3q}. \tag{45}$$

We would like to apply this approach to the shell model. However, this approach will not quite work in this case, because there is no *sweeping*, so Taylor's [45] frozen flow hypothesis is not meaningful. Moreover, there is no spatial dependence so one can only deal with time averages. Thus in order to perform a similar analysis on the shell model, we start again using equation (37) to define the time-integrated dissipation. The next ingredient is to get the connection between the shell number, $n$ and the time scale $\tau$. To an order of magnitude, in the shell model, the time derivative of $U_n$ is given as $k_n U_n^2$. Thus, there is a natural turnover time for each shell defined by

$$\tau_n(t) = |k_n U_n(t)|^{-1}. \tag{46}$$

This equation can be solved to get the dependence of $n$ upon $\tau$ at each value of $t$ and defines the function $n(\tau, t)$. The final step is to get the dissipation from the energy flux. If the shell $n$ is correctly picked so that the time-scale lies within the $n$th shell, then the dissipation should be estimated as the energy flux in that shell. In the GOY model then, the dissipation and the flux may be estimated as

$$\epsilon_\tau(t) \sim J_n(t) \sim \Im[k_n U_{n-1}(t) U_n(t) U_{n+1}(t)], \tag{47}$$

where in this equation, $n$ is to be evaluated at $n(\tau, t)$.

Because $U_n(t)$ is a fluctuating quantity, so is $\tau_n(t)$ and we cannot once and for all determine which shell belongs to which value of $\tau$. This makes the analysis of the shell model more difficult than the calculation for the real turbulence problem. To analyze the model, we must go back to the $f(\alpha)$ formalism which grew up as an alternative description of multifractal phenomena [11, 12, 46]. Here, $f(\alpha)$ is the singular spectrum as commonly defined in [12, 46, 43]. Instead of defining $\zeta_q$, we define the probability that $U_n(t)$ will have a magnitude which is of the order of $k_n^{-\alpha}$. This probability is a strongly varying function of $n$, and of the form $k_n^{f(\alpha)}$, independent of $n$. Thus, using the delta function one writes

$$\langle \delta(\ln U_n(t) + \alpha \ln k_n) \rangle \sim k_n^{f(\alpha)} \tag{48}$$

and calculates averages as integrals over $\alpha$. Thus for example,

$$1 = <1> = \int d\alpha \langle \delta(\ln U_n(t) + \alpha \ln k_n)\rangle \sim \int d\alpha \, k_n^{f(\alpha)}. \tag{49}$$

The integral is to be done by a steepest descent technique. (We only keep factors which are exponential function of $n$. Powers of $\ln k_n$ are neglected.) In this way, we learn that the maximum value of $f(\alpha)$ is zero. An analogous calculation gives a direct evaluation of $\zeta_q$. Consider

$$<|U_n(t)|^q> = \int d\alpha \langle |U_n(t)|^q \delta(\ln U_n(t) + \alpha \ln k_n)\rangle \sim \int d\alpha \, k_n^{-q\alpha + f(\alpha)}. \tag{50}$$

On the one hand this average is set by the definition of equation (26) to be $k_n^{-\zeta_q}$. On the other, the integral can be performed via steepest descents and gives the familiar Legendre transform definition of $f(\alpha)$,

$$\zeta_q = \min_\alpha [q\alpha - f(\alpha)]. \tag{51}$$

Once we know $\zeta_q$, $f(\alpha)$ is known.

Next we estimate the average of $\epsilon_\tau$ taken to various powers. Once again we compute the averages by integrating over $\alpha$. Now we have two conditions: the first being that $\epsilon_\tau$ is given by expression (47) for some appropriate value of $n$, the second being that the appropriate value of $n$ is defined via a solution of equation (46). Employing these two conditions, we find that

$$<\epsilon_\tau^q> = \int d\alpha \int dn \langle |J_n(t)|^q \delta(\ln U_n(t) + \alpha \ln k_n) \delta(\ln \tau + \ln U_n(t) k_n)\rangle. \tag{52}$$

Substituting for the flux using equation (47) and observing that the delta function permits us to do the $n$-integration, we thereby reduce the result to a simple integral over $\alpha$ of the form

$$<\epsilon_\tau^q> \sim \int d\alpha \, \tau^{X(\alpha)}, \tag{53}$$

with the exponent having the value

$$X(\alpha) = \frac{3q\alpha - 1 - f(\alpha)}{1 - \alpha}. \tag{54}$$

In the usual way, the integral is calculated by steepest descent and $\mu_q$ is determined from the saddle point integration as

$$\mu_q = \min_\alpha X(\alpha). \tag{55}$$

As desired, this equation connects the scaling exponents $\mu_q$ and $\zeta_q$. Note in particular, that $\mu_1 = 0$ holds as it should be, independently of the form of $f(\alpha)$. For similar approaches, leading to slightly different results, we refer to refs. [48].

The next step is a comparison of our theory with the results from the simulation of the model. For convenience, we denote $<\epsilon_\tau^q>$ by $\epsilon_{\tau,q}$. We expect $\epsilon_{\tau,q}$ to scale in the inertial range. It is not clear *a priori*, what the inertial range will be in the time domain, as an application of Taylor's hypothesis in the GOY model is not meaningful, see above. We find good scaling behavior for $\tau = 0.2$ to $\tau = 6$, see Figure 5 for that of $\epsilon_{\tau,2}$. To extract even preciser scaling exponents from the numerics, we plot $\epsilon_{\tau,q}$ vs. $\epsilon_{\tau,2}$ and determine the ratio $\mu_q/\mu_2$, which is a direct analogy of *extended self similarity* introduced by Benzi et. al. [47]. Figure 5 shows $\epsilon_{\tau,4}$ vs. $\epsilon_{\tau,2}$. Scaling is seen in the range between $\tau \sim 0.05$ and $\tau \sim 5$, where the times are given in the natural time units of the GOY model, the large eddy turnover times.

As stated above, to connect the $\zeta_q$ with the $\mu_q$, we need to find the singular spectrum $f(\alpha)$ from the $\zeta_q$. We take advantage of the fact that the numerical values coincide with the She-Leveque formula (32) [41]. Then $f(\alpha)$ can be easily obtained through the Legendre transformation of $\zeta_q$,

$$f(\alpha) = -2 + \frac{3(\alpha - 1/9)}{\ln(2/3)} \left( \ln\left[\frac{1/9 - \alpha}{\ln(2/3)}\right] - (1 + \ln(2/3)) \right). \tag{56}$$

We then use our formulas (53),(54) and (55) to find the ratios $\mu_q/\mu_2$. The comparison with the numerical values extracted from the scaling of $\epsilon_{\tau,q}$ versus $\epsilon_{\tau,2}$ is shown in table 2. We estimate the error in the numerical values by comparing the results of linear fits in slightly different regions of the scaling range and find it to be between 1% for small $q$ and 5% for larger $q$. Up to the $5^{th}$ moments of $\epsilon_\tau$ the agreement between the numerical results and our scaling analysis seems to be within the statistical error.

## 4.2 Time correlations for $|U_n|$

In this section, we shall estimate the scaling behavior of one of the most basic time correlation functions, namely,

$$C_{p_1 n_1, p_2 n_2}(\tau) = \langle |U_{n_1}(t)|^{p_1} |U_{n_2}(t+\tau)|^{p_2} \rangle. \tag{57}$$

We restrict ourselves to the case where $\tau$ is much larger than the typical time scales of the shells $n_1$ and $n_2$. The main idea of the calculation is to introduce a (large scale) shell $m(t)$ into (57) such, that it has a relaxation time of order $\tau$, as one can then assume, that the shells $n_1$ and $n_2$ are completely uncorrelated with $m$, whereas there should be more or less full correlation on larger scales than $m$. We thus determine the shell number $m$ through $\tau \sim \tau_m(t)$, or, by employing (46),

$$\tau \sim \tau_m(t) = \frac{1}{|U_m(t)|k_m} \sim k_m^{-(1-\beta(t))}, \tag{58}$$

with $|U_m(t)| \propto k_m^{-\beta(t)}$. We may thus write

$$C_{p_1 n_1, p_2 n_2}(\tau) = \left\langle \left|\frac{U_{n_1}(t)}{U_m(t)}\right|^{p_1} \left|\frac{U_{n_2}(t+\tau)}{U_m(t+\tau)}\right|^{p_2} |U_m(t)|^{p_1} |U_m(t+\tau)|^{p_2} \right\rangle$$

$$= \int dm \left\langle \left|\frac{U_{n_1}(t)}{U_m(t)}\right|^{p_1} \left|\frac{U_{n_2}(t+\tau)}{U_m(t+\tau)}\right|^{p_2} |U_m(t)|^{p_1}|U_m(t+\tau)|^{p_2} \right\rangle$$
$$\delta\left[(1-\beta(t))\ln k_m + \ln\tau\right]. \tag{59}$$

In terms of the shell number $m(t)$, our above restriction

$$\tau \sim \tau_m(t) \gg \tau_{n_1},\ \tau_{n_2} \tag{60}$$

reads

$$m(t) \ll n_1,\ n_2. \tag{61}$$

We thus may assume that the first two factors in (59) are independent of the last and also independent of each other due to eq. (60). On the other hand, the last two factors are assumed to be fully correlated, as $\tau \sim \tau_m$. With eq. (50) we obtain

$$C_{p_1 n_1, p_2 n_2}(\tau) \sim \int dm \left\langle \left|\frac{U_{n_1}(t)}{U_m(t)}\right|^{p_1} \right\rangle \left\langle \left|\frac{U_{n_2}(t)}{U_m(t)}\right|^{p_2} \right\rangle$$
$$\left\langle \int d\beta k_m^{f(\beta)-(p_1+p_2)\beta} \delta\left[(1-\beta(t))\ln k_m + \ln\tau\right] \right\rangle. \tag{62}$$

The shells $n_1$ and $m$ are assumed to be independent. So we only have to plug in the definition of the scaling exponents $\zeta(p_1)$, obtaining

$$\left\langle \left|\frac{U_{n_1}(t)}{U_m(t)}\right|^{p_1} \right\rangle \sim \left|\frac{k_{n_1}}{k_m}\right|^{-\zeta(p_1)}. \tag{63}$$

Doing the same for the second factor, and performing the integral over $m$ (i.e., replacing $k_m$ by $\tau^{-1/(1-\beta)}$), we find

$$C_{p_1 n_1, p_2 n_2}(\tau) \sim k_{n_1}^{-\zeta(p_1)} k_{n_2}^{-\zeta(p_2)} \int d\beta \tau^{-(f(\beta)-(p_1+p_2)\beta+\zeta(p_1)+\zeta(p_2))/(1-\beta)}. \tag{64}$$

A saddle point approximation leads to,

$$C_{p_1 n_1, p_2 n_2}(\tau) \sim k_{n_1}^{-\zeta(p_1)} k_{n_2}^{-\zeta(p_2)} \tau^{-\phi(p_1,p_2)} \tag{65}$$

with

$$\phi(p_1, p_2) = \max_\beta \frac{f(\beta) - (p_1+p_2)\beta + \zeta(p_1) + \zeta(p_2)}{1-\beta}, \tag{66}$$

which is our final result of this section. We think it is worth while to numerically check our prediction (66) within the GOY dynamics and possibly try to extend it to Navier-Stokes dynamics.

## 4.3 Lyapunov indices and correlation functions

As one application of the correlation function analysis of the previous subsection, we turn to an analysis of the largest Lyapunov index of the GOY model, called $\lambda_L$. Our work follows closely the approach of A. Crisanti, M.H. Jensen, G. Paladin, and A. Vulpiani [49], hereafter called CJPV.

By defining $\delta u$ the infinitesimal difference between two (vector) fields evolving under the same equation (i.e. the GOY model in our case), we can define the exponential divergence rate $\gamma$ after a time delay $\tau$ of two trajectories close at time $t$:

$$\gamma_\tau(t) = \frac{1}{\tau} ln \frac{||\delta u(t+\tau)||}{||\delta u(t)||}. \tag{67}$$

Using (67) we can define $\lambda_L$ as the limit for infinite $\tau$ of $\gamma_\tau(t)$ and $\gamma_0(t)$ or simply $\gamma(t)$ as the local divergence rate for trajectories.

CJPV assumed that the Lyapunov exponent should be proportional to the average of the largest characteristic rate of change of the system, namely the inverse of the eddy turnover time $\tau_d$ at the dissipation scale [49]. In the GOY model, this rate is

$$\gamma = u_n k_n \simeq k_n^{1-\alpha}. \tag{68}$$

We define $n$ to correspond to a dissipation scale by equating the viscous term $D_n \sim \nu k_n^2 U_n$ and the nonlinear term $C_n \sim k_n U_n^2$ of the GOY equation, giving

$$U_{n(t)}/k_{n(t)} = \nu \sim Re^{-1}. \tag{69}$$

Here we have called $\nu$ the inverse Reynolds number $Re$ of the GOY dynamical equations. From (68) and (69) we obtain

$$\lambda_L = <\gamma(t)> = \int k_n^{1-\alpha+f(\alpha)} \delta[(1+\alpha)(\ln k_n) - \ln Re] d\alpha dn \simeq Re^a, \tag{70}$$

where

$$a = \max_\alpha \frac{f(\alpha) + 1 - \alpha}{1 + \alpha}. \tag{71}$$

The numerical value of $a$ estimated by CJPV is 0.46. With $f(\alpha)$ from (56) we obtain $a = 0.47$. K41 predicts $a = 0.5$. It follows that the exponent $a$ is not strongly effected by the intermittency of the system.

The next step is to calculate the variance in the Lyapunov indices, which CJPV estimated by numerically evaluating the integral

$$\mu_L \simeq \int_0^\infty \langle [\gamma_0(t+s) - \lambda_L][\gamma_0(t) - \lambda_L] \rangle ds \tag{72}$$

which measures the strength of fluctuations in the system. In particular it has been shown [36] that $\frac{\mu_L}{\lambda_L} = 1$ separates weak from strong intermittency. The numerical computations [49] show that $\mu_L \simeq Re^w$ where $w = 0.8$.

To estimate the integral in equation(72) one notices that the integrand involves a correlation of two factors $|k_n U_n|$ at different times. This is just the kind of integral estimated in the previous subsection, except that $n$ is fixed to lie at the dissipation scale. Instead of redoing the same calculations it is instructive to use an order of magnitude argument to estimate $\mu_L$.

When the time separation in equation(72) is of order of the integral scale-value, $s \sim 1$, the fluctuations in $\gamma(t)$ are also of relative order unity. This result is a consequence of the fact that $\gamma(t)$ is proportional to a velocity, and the velocity is a product of roughly independent random variables, one for each shell. Each variable has fluctuations of order unity, and the ones for the first shell in $\gamma(t)$ and $\gamma(t+s)$ are only weakly correlated when $s$ is of order unity. Thus the fluctuations in the integrand are of order of the uncorrelated part, so that this part of the integral gives

$$\mu_L \sim \int_{\tau_d}^{1} \lambda_L^2 ds \sim \lambda_L^2, \tag{73}$$

where $\tau_d$ is the dissipative time scale. For $s \ll 1$ there is of course no contribution. To estimate the remaining part $\int_0^{\tau_d}$ of the integral one goes back to equation(65) and imagine integrating a result like this over $\tau$. If $\phi$ is greater than one, the integral will contribute for small $\tau$; otherwise, the main contribution will occur for $\tau$ of order unity. But, using the formula following from She and Leveque, equation (56), one can see that $\phi$ is far smaller than one. Hence the main contributions comes at the integral scale and we can argue

$$\mu_L \sim \lambda_L^2. \tag{74}$$

For the standard GOY model parameters $\epsilon = 0.5$, $\lambda = 2$ this gives us an estimate $\mu_L \sim Re^{2 \cdot 0.47} = Re^{0.94}$ which has to be compared with the numerical result of CJPV, $\mu_L \sim Re^{0.8}$. Considering the numerical uncertainty and the lack of rigor of our order of magnitude argument, the agreement is not too bad.

We also have to remind that the scaling properties of $\lambda_L$ are linked to small fluctuations of the multiplicative process, i.e. in the region where the SL formula disagrees with our numerical findings.

The computations discussed so far make the (strong) assumption that the instantaneous Lyapunov exponent is controlled by the instantaneous velocity field at the dissipation time scale. In fact, between the two quantities there might be a time lag $\tau_R$ which characterizes the relaxation time scale for the instantaneous Lyapunov exponent to become proportional to the inverse dissipation time scale. $\tau_R$ is not simply linked to the scaling properties of the GOY model and its multifractal behavior. If $\tau_R$ is small compared to the dissipation time scale, then the analysis previously described gives the expected scaling of the intermittent exponent $\mu$ with respect to the Reynolds number. On the other hand, for large enough $\tau_R$ one should find a smaller intermittency effect in the system.

These examples give us some feeling that we might have a crude but useful understanding of time dependence in the GOY model. Yet more numerical and analytical analyses is definitely necessary.

# 5 Conclusions

The dynamical equations of the GOY model permit a solution of a type described by BBP in which there are roughly independent fluctuations in all of the shells of the inertial range. The long-ranged structure of these fluctuations give all the multifractal scaling properties. The conservation laws for energy and helicity play an essential role in the structure of the solution, but such 'details' as the momentum scale $\lambda$, the Reynolds number, and the form of the viscous cutoff might also determine the exact scaling exponents.

All scaling that we have examined in the model seems to be determined by one set of multifractal exponents, e.g. those of the velocity. In particular, many time-dependent correlations may be estimated from these exponents. However, at the moment there exists no accurate way of calculating these scaling exponents except by direct numerical simulation. However, see [41] and [50] for an interesting set of insights into the possible structure of the multifractal behavior.

This paper devotes some attention to the scaling of time correlations. The corresponding frequency spectra have recently been studied in [51]. Notice that these are the quantities which are measured in experimental studies of turbulence. Turbulence theory up to now has mainly focused on spatial structure functions and wave vector spectra and has connected them with time structure functions and frequency spectra only via Taylor's hypothesis. Clearly time-dependence is worth studying in its own right.

We finally stress, that the exact way multifractality works itself out in the GOY dynamics is slightly different from what happens in real turbulence because there is no analog of the sweeping which plays such an essential role in Navier- Stokes dynamics, and, of course, because the GOY equations are only a model which might or might not catch essential features of the Navier-Stokes equations.

**Acknowledgements:** We particularly appreciate many useful exchange with S. Esipov who has helped us in many ways. Z.S. She has shared his ideas with the Chicago part of our group and changed some of the thinking about the relevance of the form of the dissipation. We thank T. Dupont for providing us the source of lsode solver and the HLRZ Jülich for supplying us with computer time. It is our pleasure to thank J. Eggers, O. Gat, S. Grossmann, M. Jensen, I. Procaccia, N. Schörghofer, and R. Zeitak for useful exchange of information. D.L. acknowledges support by a NATO grant through the Deutsche Akademische Austauschdienst (DAAD). This research was supported in part by the DOE, the ONR, and by the MRSEC Program of the National Science Foundation under Award Number DMR-9400379.

\* On leave of absence from Fachbereich Physik, Universität Marburg, Renthof 6, D-35032 Marburg.

# Tables

| n | numerics $\epsilon = 0.3$ | eq. (25) $\epsilon = 0.3$ | numerics $\epsilon = 0.5$ | eq. (25) $\epsilon = 0.5$ |
|---|---|---|---|---|
| 4 | 0.15000 | 0.15000 | 0.14 | 0.14 |
| 5 | 1.31664 | 1.31667 | 1.35 | 1.10 |
| 6 | 0.28291 | 0.28291 | 0.27 | 0.36 |
| 7 | 0.76848 | 0.76857 | 0.78 | 0.59 |
| 8 | 0.37766 | 0.37770 | 0.37 | 0.46 |
| 9 | 0.61321 | 0.61333 | 0.62 | 0.52 |
| 10 | 0.43483 | 0.43532 | 0.43 | 0.48 |
| 11 | 0.55055 | 0.55110 | 0.55 | 0.50 |

Table 1: The $r_n$ from our numerical calculation, compared with the expression (25) when $\epsilon = 0.3$ and $\epsilon = 0.5$. In the latter case, we show the time averaged velocities. Note that the same alternation between high and low values occurs in all four columns.

| q | derived from $\zeta_q$ via (55) | direct numerical result | derived from (32) |
|---|---|---|---|
| 1.0 | 0.00 | $0.0005 \pm 0.0002$ | 0.00 |
| 1.5 | 0.41 | $0.42 \pm 0.03$ | 0.40 |
| 2.0 | 1.00 | $1.00 \pm 0.03$ | 1.00 |
| 2.5 | 1.72 | $1.71 \pm 0.04$ | 1.77 |
| 3.0 | 2.54 | $2.52 \pm 0.1$ | 2.67 |
| 3.5 | 3.44 | $3.41 \pm 0.2$ | 3.68 |
| 4.0 | 4.42 | $4.25 \pm 0.2$ | 4.78 |
| 4.5 | 5.44 | $5.23 \pm 0.4$ | 5.95 |

Table 2: The ratio $\mu_q/\mu_2$ from our data for velocity scaling compared with the ratio deduced from the the scaling analysis of dissipation. The numerical errors are estimated from the differences between the fits in slightly different fitting regions. The error in the first column due to the numerical error in the $\zeta_q$ is much less significant compared to the error in the second column. In the last column we give $\mu_q/\mu_2$ as it follows from eqs. (32) and (45). These values are systematically slightly larger than the first two columns, yet the agreement is satisfactory.

**Figure captions**

Figure 1: Elimination of period three oscillations. We show three curves plotted against shell number $n$. Two are for $s_{n,1} = \langle |U_n| \rangle$: A plot for the normal viscosity case ($D_n = \nu k_n^\phi U_n$ with $\phi = 2$) and another for the hyperviscosity case ($\phi = 4$). Both of these show period-three oscillations. The remaining curve plots $S_{n,1}$ for the hyperviscosity case and shows no period-three oscillations. In each case time averages are taken over 15 000 large eddy turnover times, the number of shells is 22. One curve is for our standard model parameters, the hyperviscosity cases have $\nu = 10^{-15}$.

Figure 2: Different kinds of scaling analyses with increasing accuracy: $s$, which is the magnitude of the velocity, $S$ which is a cube root of the imaginary part of a product of three velocities, and $\Sigma$, which is the cube root of the energy flux. Part a shows curves drawn for $q = 1$ ; part b for $q = 6$. In both cases, $\Sigma$ gives the longest scaling range, and hence probably the best estimates for scaling exponents. All curves are drawn for the standard parameter values. For comparison, K41 scaling is also shown.

Figure 3: The deviations $\delta\zeta_q$ from Kolmogorov scaling are compared with one another for different parameter sets. We found it most instructive to plot $\delta\zeta_q/(q(q-3))$ against $q$, to eliminate the trivial agreement at $q = 0$ and $q = 3$. The dashed line is our numerical calculation, with the standard parameter values. The solid line, which is impressively close to our numerical results for $q > 3$, is the result (32) of She and Leveque [41]. The dotted line reports the shell results for another set of parameters, namely $\epsilon = 0.7$ and $\lambda = 2$. Note the pronounced disagreement with the SL result in this case.

Figure 4: $\delta\zeta_q$ versus $q$ for four sets of parameter values. Three parameter pairs $(\epsilon, \lambda)$ lie on the curve (33) which defines the right value of the helicity. These have $\lambda = 10/3$ and $\lambda = 10/7$ paired with their corresponding $\epsilon$'s. The last value lies off the curve and has $\epsilon = 0.7$ with $\lambda = 2$. Note how the values on the curve stand grouped together in comparison with the other one.

Figure 5: Scaling corrections $\delta\zeta_q$ for three different forms of viscous damping. In each case, the damping term is $D_n = \nu k_n^\phi U_n$. The solid line shows the result for $\phi = 2$ and $\nu = 10^{-7}$, the dashed line for $\phi = 4$ with $\nu = 10^{-15}$, and the dotted line for $\phi = 6$ with $\nu = 5 \cdot 10^{-25}$.

Figure 6: The top picture shows the scaling behavior of $\epsilon_{\tau,2}$ in $\tau$, while the bottom employs the idea of extended-self similarity to plot one moment against another, i.e. $\epsilon_{\tau,4}$ against $\epsilon_{\tau,2}$. The slope of the best fit is also shown (dashed line).

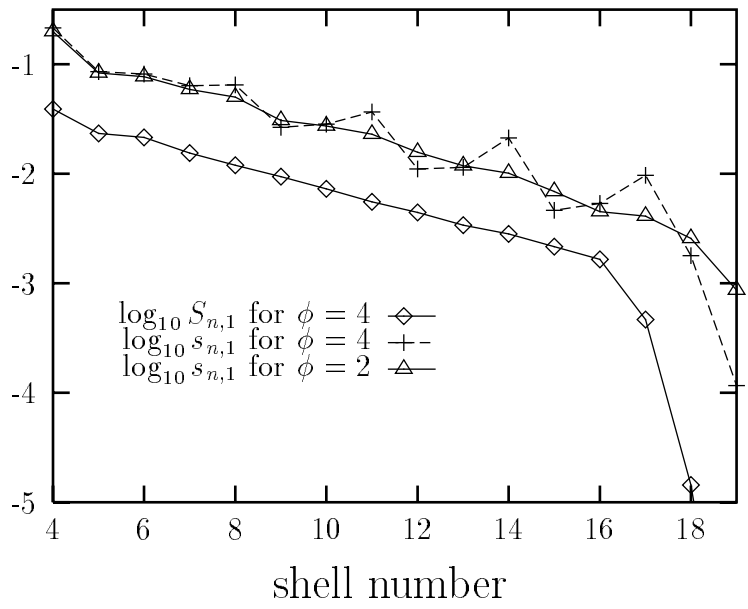

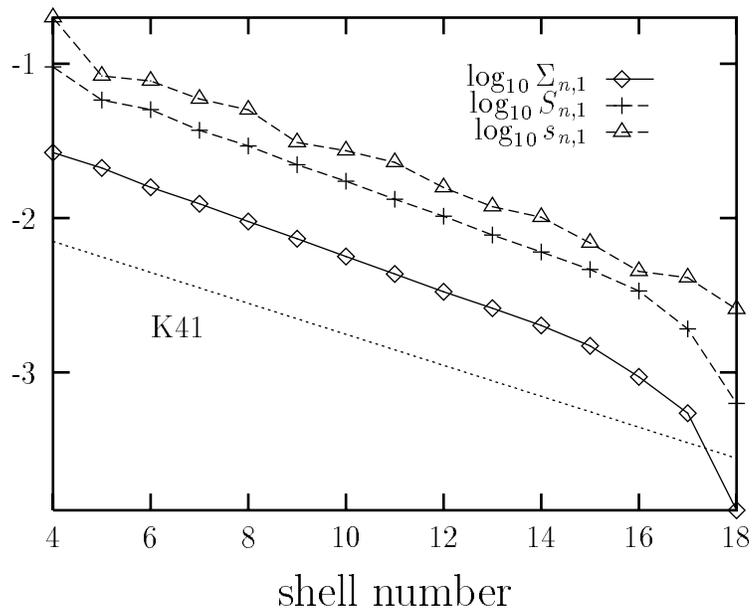

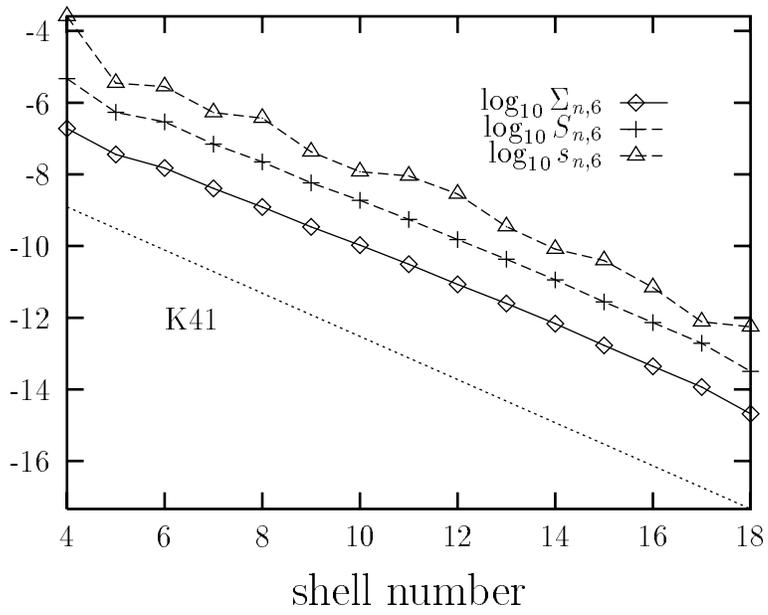

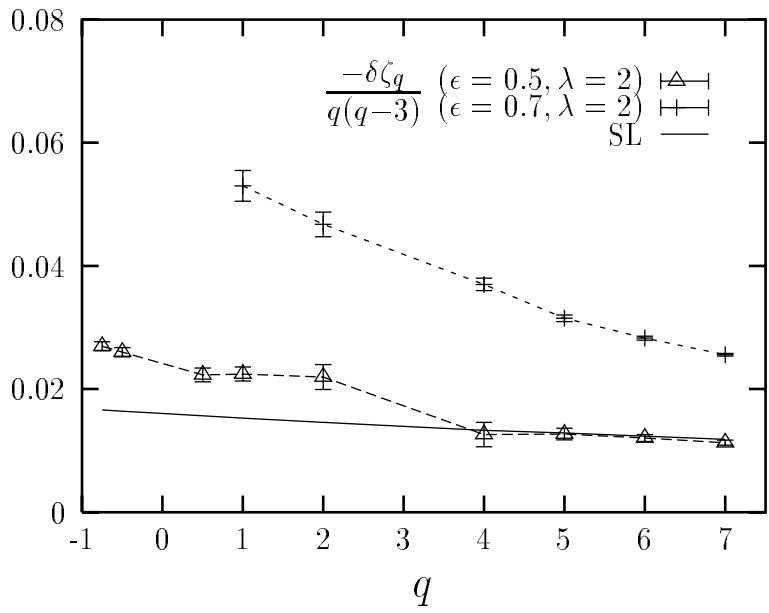

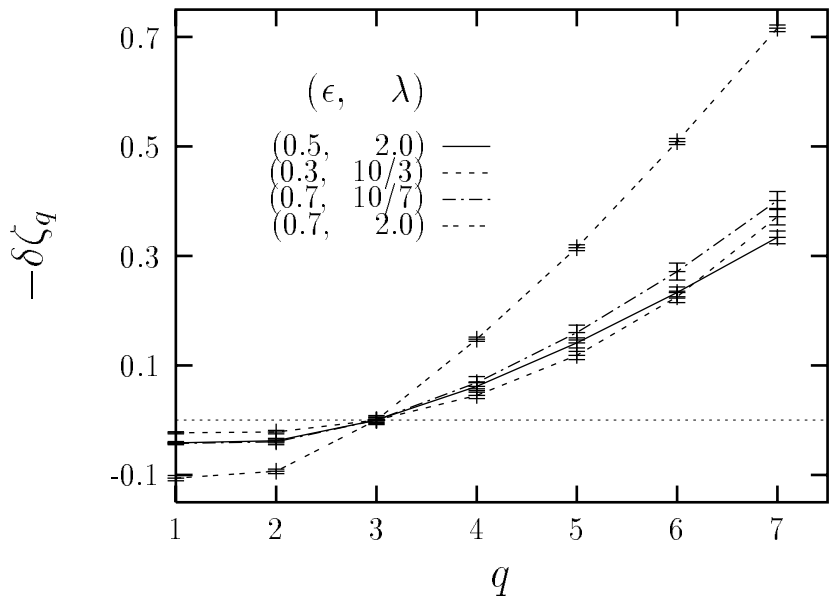

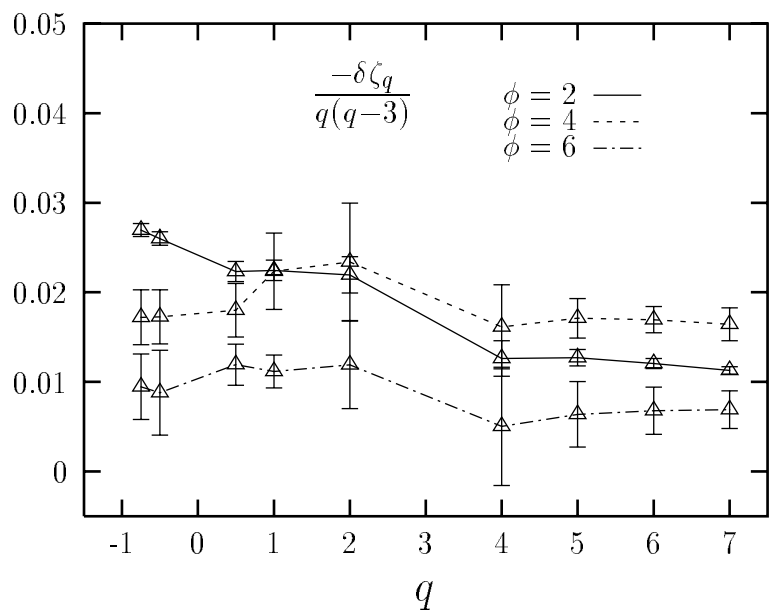

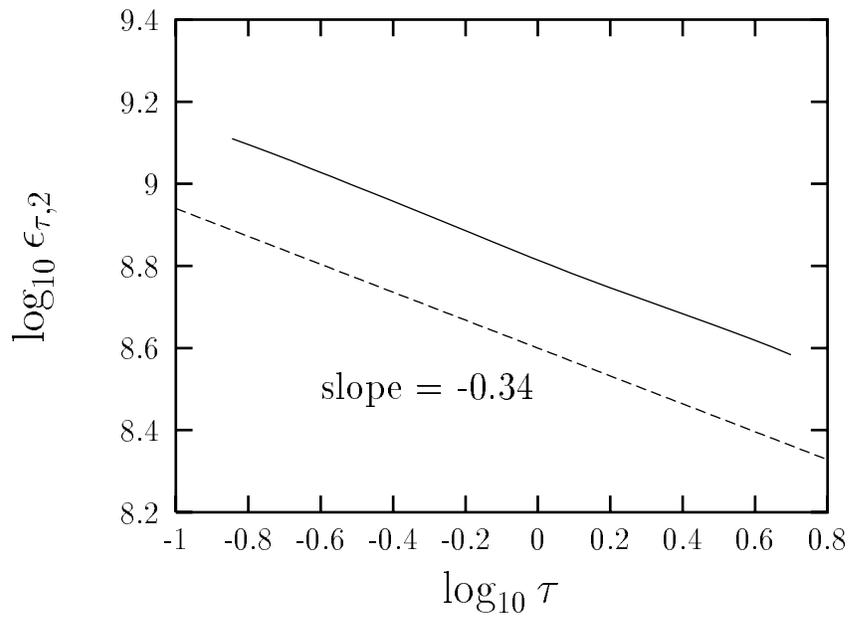

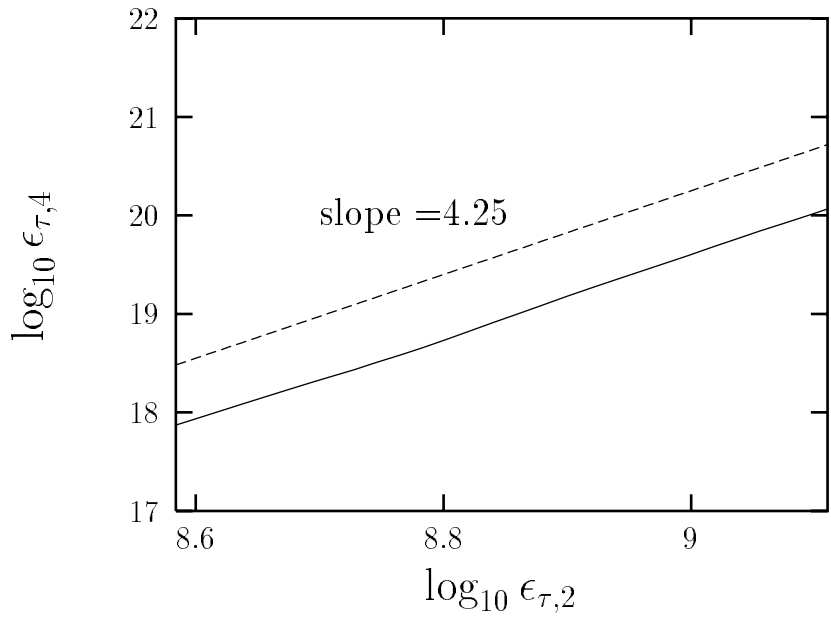